\documentstyle[aaspp4,flushrt,psfig]{article}
\begin{document}

\title{Inhomogeneous Absorbers and Derived Column Densities}
\author{Bassem M. Sabra}
\affil{American University of Science and Technology, P. O. Box 95, Zahle, 
LEBANON} 
\author{Fred Hamann}
\affil{Department of Astronomy, University of Florida, Gainesville, 
FL 32611, USA}

\affil{\it{ApJ, submitted}}

\begin{abstract}
We study the dependence of column densities derived from 
absorption lines on the spatial distribution of the 
ions in an absorber. In particular, we investigate four 
varieties of coverage by the absorber of the background 
source: the familiar homogeneous partial coverage (HPC), and three 
functional forms that parameterize inhomogeneous coverage: 
a powerlaw, an ellipse quandrant, and a Gaussian distribution.  
We calculate the residual line intensities obtained from our 
inhomogeneous coverage models and then use these intensities as 
``observed'' quantities to compute the optical depth and covering 
factors assuming HPC.
We find that the resulting spatially-averaged optical depths 
are comparable (within a factor of $\lesssim 1.5$) to the average 
optical depths of the input distributions, as long as the input 
distributions do not contain spatially narrow ``spikes.'' Such spikes 
(very large optical depths over a small coverage area) can profoundly 
affect the average optical depth in the absorber but have little 
impact on the observed intensities. 
We also study the converse approach: we start with HPC 
as the assumed physical model and then infer the parameters of the 
inhomogeneous 
coverage models through a doublet analysis. Again the 
resulting average optical depths are comparable to the 
corresponding quantity of the input distribution. 
Finally, we construct a more realistic two-dimensional optical 
depth distriubution based on a random distribution of absorbing 
clouds, and we use that to calculate observed intensities. A 
doublet analysis applied to those intensities shows all four of 
our simple analytic functions yield an accurate estimate of the 
true average optical depth, while the powerlaw yields the best 
approximation to the intial distribution.

\end{abstract}


\section{Introduction}
Resonant absorption lines observed in the spectra of many astronomical 
objects provide important diagnostics about the physical and chemical states of 
the environments in which they originate. 
One of the important aims of absorption line studies is determining accurate elemental 
abundances. The steps involved in using absorption lines as abundance diagnostics are the 
following: Derive ionic column 
densities in a way that accounts for an absorber that might be inhomogeneous and/or 
partially covering the background light source, and convert column density ratios to 
abundance ratios using ionization corrections. In this paper we are going to discuss the 
first issue, namely the effect of the spatial distribution of ions on the column densities 
as derived from the observed absorption lines.  
 
A common implicit assumption made when studying absorption line systems 
in general 
is that a compact source, such as a star or a quasar, irradiates a cloud whose 
internal structures are large compared to the projected size of the continuum 
source. However, in the more general case, we must consider the 
two-dimensional distribution of optical depths, $\tau(x,y,\lambda)$, across 
the projected area of the emitting source. 
The observed residual intensity in an absorption line is then:\\
\begin{equation}
I(\lambda)=\int\int S(x, y, \lambda)~e^{-\tau(x,y,\lambda)}~\frac{dxdy}{A}, 
\end{equation}
where $S(x, y, \lambda)$ is the two-dimensional intensity distribution, 
emitted by the source at wavelength 
$\lambda$, $A=\int\int dxdy$ is the total projected area of the source, 
and $\tau(x, y, \lambda)$ is the optical depth distribution across 
the surface of the continuum emitter. The spatial distribution of the 
column density per unit wavelength $\lambda$, $N_\lambda (x,y)$, is related to 
$\tau(x,y,\lambda)$ by:
\begin{equation}
N_\lambda(x, y) d\lambda=\frac{m_e c^2}{\pi e^2 f \lambda^2} \tau(x,y,\lambda) d\lambda,
\end{equation}
where $m_e$ is the mass of the electron, $c$ is the speed of light, 
$e$ is the charge of the electron, 
and $f$ is the oscillator strength of the line. 
Hereafter, we will always assume dependence 
on wavelength, though we will not always indicate it, for the sake of 
not cluttering the equations below. 

Equation (1) expresses the most general situation where the observed 
spectrum is an intensity-weighted average of $e^{-\tau(x, y, \lambda)}$ 
over the projected area of the emitter. The residual intensity in an absorption 
line depends non-trivially on the optical depth distribution $\tau(x, y, \lambda)$. 
This paper examines a wide range of possible spatial optical depth distributions, 
characterized for convenience by three functional forms, to study 
the effects of the shape of the optical depth distribution on the absorption 
lines, and conversely, on our ability to use the observed absorption lines 
as diagnostics of the optical depths and column densities in the 
absorber. 

The ionic column density 
derived from absorption lines depends on how the absorbing 
cloud covers the continuum source: HPC 
(Wampler et al. 1995; Barlow, Sargent, \& Hamann 1997; 
Hamann et al. 1997), or inhomogeneous coverage 
(de Kool, Korista, \& Arav 2002). 
The term {\it homogeneous} as employed in this paper will always refer to the 
optical depth being constant across all or part of 
the spatially extended emission source. We 
extend the formalism introduced in de Kool et al. (2002), which we will 
refer to as inhomogeneous partial coverage (IPC), to explore the effects 
of different variants of inhomogeneous coverage and to compare them with HPC 
over a wide range of physical conditions. Since we do not know what 
the optical depth distributions are in real absorbers, it is important to 
consider a range of possibilities and compare their effects on line strengths 
and derived column densities. 
We show how to reparameterize a two-dimensional 
optical depth distribution to a one-dimensional function in \S 5 and show that 
the simple IPC prescriptions we study here provide relatively accurate 
representations of more complicated distributions. 

\subsection{HPC Analysis}
In the spectra of quasar absorption line systems, for example, it has 
been pointed out that doublet ratios sometimes 
differ from what one would expect from atomic physics
(e.g., Petitjean, Rauch, \& Carswell 1994; Wampler, Chugai, \& Petitjean 
1995; Barlow et al. 1997; Hamann et al. 1997). 
Hamann et al. (1997) provided an explanation of the 
ratio discrepancy in terms of a HPC model. 
In this model, the absorbing cloud covers only a fraction, $C_f$, of the 
continuum source, with the implicit assumption that the coverage 
fraction for both lines is the same (e.g., see Ganguly et al. 1999; 
Gabel et al. 2002 for situations when this is not the case).  
The residual intensities in the doublet lines according 
to this HPC model are a special case of Eq. (1), normalized to 
unity in the continuum:
\begin{equation}
I_s=(1-C_f)+C_f e^{-\tau_s}
\end{equation}
for the stronger line, and 
\begin{equation}
I_w=(1-C_f)+C_f e^{-\tau_w}
\end{equation}
for the weaker line. The $C_f$'s and the $\tau$'s are also functions 
of $\lambda$. Given $I_s$ and $I_w$, at each wavelength across the 
profile, for a particular 
doublet, one can solve simultaneously for the coverage fraction, $C_f$, 
and $\tau_w = \frac{f_w~\lambda_w}{f_s~\lambda_s}~\tau_s$, where 
$f_s$ and $f_w$ are the 
oscillator strengths of the strong and weak lines, respectively, 
while $\lambda_s$ and $\lambda_w$ are wavelengths of the line doublet.  
Ganguly et al. (1999) gave a general expression for $C_f$ when 
$\tau_s : \tau_w$  is not necessarily $2 : 1$. For this 
special $2 : 1$ case, such as for 
\ion{C}{4} $\lambda\lambda 1548, 1551$, Hamann et al. (1997) and 
Barlow et al. (1997) found that:
\begin{equation}
C_f=\frac{I_w^2-2~I_w+1}{I_s-2~I_w+1},
\end{equation} 
and the optical depth in the stronger line of the doublet is:
\begin{equation}
\tau_s=2 \ln\left(\frac{1-I_w}{I_w-I_s}\right). 
\end{equation}
 
The HPC model uncovered additional layers of complexity in dealing 
with absorption lines. Barlow et al. (1997) found evidence that 
the coverage fraction can vary with velocity across the absorption 
profiles. Arav et al. (1999) discovered an apparent trend between the 
coverage fraction and the ionization potential of the absorbing 
ion in the Broad Absorption Line (BAL) QSO PG~0946+301. Another apparent 
trend was found by Hamann et al. (2001) in which stronger lines 
seem to have higher coverage fractions. The HPC analysis prompted 
de Kool et al. (2002) to introduce an inhomogeneous   
coverage prescription (the IPC, we give a detailed description below).  
The important difference between the HPC and IPC is the inhomogeniety 
of coverage. In the IPC formalism the absorbing cloud partially, or sometimes 
completely, covers the continuum source, but inhomogeneously, i.e. 
at some locations the optical depth is high, while at others it is 
low or simply absent. de Kool et al. (2002) parameterized the spatial 
distribution of optical depths across the surface of the source as a 
powerlaw in one dimension. Our study will investigate this powerlaw and 
other parameterizations, provide extensive comparisons with the HPC analysis 
over a wide range of optical depths, and use the IPC prescriptions 
directly to derive optical depths from residual intensities synthesized 
using Eqs. (3) and (4), and develop a formalism to derive properties of 
the IPC distributions based on the ratios of absorption lines in doublets.

\section{The IPC Models}
A natural question arises in light of the IPC models: 
What is the effect of the shape of the optical depth distribution 
on the ionic column densities one measures from the absorption lines, 
and how do these column densities compare with those derived under the 
HPC assumption? To address these issues, we first show how an  
arbitrary optical depth distribution determines the residual intensity 
level in an absorption line (based on de Kool et al. 2002). 

Consider a two-dimensional continuum source with a intensity, $S(x, y, \lambda)$, 
covered with a cloud with an optical depth distribution $\tau(x,y,\lambda)$. 
The resulting residual intensity $I(\lambda)$ is given by Eq. (1). If the 
average emitted intensity is:\\
\begin{equation}
S_0(\lambda)=\int\int S(x, y, \lambda) \frac{dxdy}{A},  
\end{equation}
then we can define $f(\tau, \lambda)d\tau$, the fraction of the 
intensity covered by optical depths between $\tau$ and $\tau+d\tau$, as:
\begin{equation}
f(\tau, \lambda)d\tau=\frac{S(x, y, \lambda)}{S_0(\lambda)} \frac{dxdy}{A}. 
\end{equation}
If the source has uniform brightness, then the fraction of the area 
covered by optical depths between $\tau$ and $\tau+d\tau$ is:
\begin{equation}
\frac{dxdy}{A}=f(\tau, \lambda)d\tau. 
\end{equation}
With these definitions Eq. (1) reduces to, in normalized units ($A=1$) and $S_0=1$:
\begin{equation}
I(\lambda)=\int_{\tau_{min}}^{\tau_{max}} f(\tau,\lambda)~e^{-\tau(\lambda)}~d\tau,
\end{equation}
where $\tau_{max}$ and $\tau_{min}$ are the maximum and minimum optical 
depths, respectively, at $\lambda$. Without 
loss of generality, we can rearrange any two-dimensional 
optical depth distribution to a one-dimensional distribution (see \S 5) 
that preserves the fraction of the source (intensity or area) covered at 
each wavelength. With all these simplifications, the fraction 
of the area covered by optical depths between $\tau$ and $\tau+d\tau$ 
is $dx=f(\tau, \lambda)d\tau$, where $x$ is a normalized spatial 
coordinate between 0 and 1.   

We consider here three examples for the optical depth distribution. 
In the first one, the optical depth, or equivalently the column density 
of some ion, follows a powerlaw distribution in the spatial coordinate 
$x$, as in de Kool et al. (2002). In the second parameterization, the 
optical depth distribution looks like one quadrant of an ellipse 
The third parametrization is a Gaussian distribution. These distribution 
functions represent an extremely wide range of possible optical depth 
distributions. The powerlaw and Gaussian are different variations 
on inhomogeneous complete coverage, while the Ellipse is similar to HPC 
but with the inclusion of inhomogeneous 
{\it partial} coverage (see \S 2.2). It is worth keeping in mind that HPC is a special 
case of the IPC analysis where the spatial optical depth distribution 
involves a step function, a fact which leads to the relatively simple 
Equations (3) and (4). Our study will go beyond the results presented 
in de Kool et al. (2002) not only by investigating the Gaussian and Ellipse 
parameterizations, in addition to the powerlaw, but also by presenting a coherent, 
easy-to-follow formalism to glean information from absorption line data.  
The treatment of the powerlaw in this paper will be more general than that in 
de Kool et al. (2002), in which $\tau_{min}$ was restricted to zero. 

\subsection{Powerlaw Distribution}
With the assumptions taken above, an optical depth described by a powerlaw 
distribution is then mathematically expressed as: 
\begin{equation}
\tau=(\tau_{max}-\tau_{min}) x^a+\tau_{min},
\end{equation}
where $\tau_{max}$ and $\tau_{min}$ are the maximum and minimum 
optical depths, respectively, $a$ is the powerlaw index, and $x$ 
is a normalized spatial coordinate between 0 and 1. 
We show examples in Figure 1. A value of 
$a=1$ describes a straight line. The powerlaw index in Eq. (11) is 
related to the $p$ parameter in de Kool et al. (2002) by $a=1/(p+1)$.  
For our powerlaw distribution, the fraction of the area covered by 
optical depths between $\tau$ and $\tau+d\tau$ is:
\begin{equation}
dx=\frac{1}{a} \frac{(\tau-\tau_{min})^{(1-a)/a}}{(\tau_{max}-\tau_{min})^{1/a}}~d\tau,  
\end{equation}
which corresponds to:
\begin{equation}
f(\tau,\lambda)=\frac{1}{a} \frac{(\tau-\tau_{min})^{(1-a)/a}}{(\tau_{max}-\tau_{min})^{1/a}}. 
\end{equation}
To calculate the residual intensity, we substitute Eq. (13) in Eq. (10) and 
carry out the integration. The result is: 
\begin{equation}
I(\lambda)=\frac{1}{a} \frac{e^{-\tau_{min}}}{(\tau_{max}-\tau_{min})^{1/a}} \Gamma(1/a)P(1/a, \tau_{max}-\tau_{min}), 
\end{equation}
where $\Gamma$ and $P$ are the complete and incomplete Gamma functions, 
respectively. It is important to keep in mind that, in principle, there 
exist different values of $\tau_{max}$, $\tau_{min}$, and $a$ 
at every $\lambda$. Given a specific optical depth spatial 
distribution (Eq. 11), Eq. (14) gives the residual intensity at each 
wavelength across an observed absorption line.   

\subsection{Ellipse Distribution}
We will refer to the second optical depth spatial distribution as 
the ``Ellipse''. It can be expressed mathematically as:
\begin{equation}
\frac{x^2}{b^2}+\frac{(\tau-\tau_{min})^2}{(\tau_{max}-\tau_{min})^2}=1, 
\end{equation}
where $\tau_{max}$ ($\tau_{min}$) are the maximum (minimum) optical depths with
($\tau_{max}-\tau_{min}$) being the semi-major axis of the ellipse, while 
$b$ is its the semi-minor axis. We show in Figure 2 examples 
for three different $b$ values. As before, the optical depths and $b$ are, in principle, 
functions 
of wavelength. The Ellipse is similar to the HPC model 
if $b<1$, with $b$ playing a role similar to that 
of $C_f$. The fraction of the area covered by optical depths 
between $\tau$ and $\tau+d\tau$ is:
\begin{equation}
dx=-\frac{b}{(\tau_{max}-\tau_{min})}\frac{(\tau-\tau_{min})d\tau}{\sqrt{(\tau_{max}-\tau_{min})^2-(\tau-\tau_{min})^2}},
\end{equation}
where the minus sign only indicates that the 
optical depth is decreasing with increasing $x$ (see Figure 2). With this we have:  
\begin{equation}
f(\tau,\lambda)=\frac{b}{(\tau_{max}-\tau_{min})}\frac{(\tau-\tau_{min})}{\sqrt{(\tau_{max}-\tau_{min})^2-(\tau-\tau_{min})^2}}. 
\end{equation}
Following the procedure outlined above, the residual intensity 
for $b<1$ is:
\begin{equation}
I(\lambda)=(1-b)+\frac{b}{(\tau_{max}-\tau_{min})} \int_{\tau_{min}}^{\tau_{max}} \frac{(\tau-\tau_{min})~e^{-\tau}}{\sqrt{(\tau_{max}-\tau_{min})^2-(\tau-\tau_{min})^2}}~d\tau.
\end{equation}
For $b \ge 1$, we have: 
\begin{equation}
I(\lambda)=\frac{b}{(\tau_{max}-\tau_{min})} \int_{\tau_{m}}^{\tau_{max}} \frac{(\tau-\tau_{min})~e^{-\tau}}{\sqrt{(\tau_{max}-\tau_{min})^2-(\tau-\tau_{min})^2}}~d\tau, 
\end{equation} 
where $\tau_m=\tau_{min}+\frac{(\tau_{max}-\tau_{min})}{b} \sqrt{b^2-1}$. 
For the $b \ge 1$ case, $\tau_m$ (not $\tau_{min}$) is the actual 
minimum optical depth realized in the absorber (the same also applies 
for the Gaussian distribution below). These integrals will 
have to be solved numerically at every $\lambda$ given $b$, $\tau_{max}$, 
and $\tau_{min}$. 

\subsection{Gaussian Distribution}
We use a Gaussian as a third example of an optical depth distribution:   
\begin{equation}
\tau=(\tau_{max}-\tau_{min})~e^{-\frac{x^2}{2\sigma^2}}+\tau_{min}, 
\end{equation}
where $\sigma$ is the standard deviation of the Gaussian function 
(Zombeck 1990). This functional form can describe cases of a small 
percentage of the maximum optical depth covering a large fraction 
of the projected area of the continuum source. For instance, 
if $\sigma \lesssim 1/3$ and $\tau_{min}=0$ then 
$\tau \lesssim 0.01\tau_{max}$ over $\gtrsim (1-3\sigma)$ 
the projected area of the emitter. 
Distributions for three values of $\sigma$ are shown in Fig. 3.
The covered area is: 
\begin{equation}
dx=-\frac{\sigma}{\sqrt{2}}\frac{d\tau}{(\tau-\tau_{min})\sqrt{\ln\frac{\tau_{max}-\tau_{min}}{\tau-\tau_{min}}}}, 
\end{equation}
where the minus sign is merely due to the 
arbitrary parameterization (see Figure 3). This area fraction corresponds to:
\begin{equation}
f(\tau, \lambda)=\frac{\sigma}{\sqrt{2}}\frac{1}{(\tau-\tau_{min})\sqrt{\ln\frac{\tau_{max}-\tau_{min}}{\tau-\tau_{min}}}}. 
\end{equation}
Using the procedure outlined above, we evaluate the residual intensity:
\begin{equation}
I(\lambda)=\frac{\sigma}{\sqrt 2}~\int_{\tau_m}^{\tau_{max}} \frac{e^{-\tau}~d\tau}{(\tau-\tau_{min})~\sqrt{\ln~\frac{\tau_{max}-\tau_{min}}{\tau-\tau_{min}}}},
\end{equation}
where $\tau_m=(\tau_{max}-\tau_{min})~e^{-1/2\sigma^2}+\tau_{min}$. 
It is important to keep in mind that $\tau_m$ is the minimum 
optical depth achieved by the this distribution. 
Given $\sigma, \tau_{max}$, and $\tau_{min}$ , we can numerically  
compute the integral in Eq. (23) to calculate $I(\lambda)$.

\section{Comparisons}
In this section, we compare the IPC models to the HPC case as follows. 
We compute the absorption line intensities and spatially averaged optical 
depths for different values of $\tau_{min}$, $\tau_{max}$ and the 
shape parameters, $a$, $b$, and $\sigma$, for each of the three IPC 
distribution functions described above. We use intensities computed 
for doublets having a $2 : 1$ optical depth ratio to derive 
the coverage fraction and optical depth assuming HPC, based on  
Equations $3-6$ in \S1. The spatially-averaged optical depth, 
$\tau_{avg}$, for the stronger line in the doublet is:
\begin{equation}
\tau_{avg}=\frac{\int^1_0 \tau~dx}{\int^1_0 dx}=\int^1_0 \tau~dx.
\end{equation} 
For the powerlaw distribution, this quantity is:
\begin{equation}
\tau_{avg}^{Powerlaw}=\frac{\tau_{max}+a~\tau_{min}}{1+a},
\end{equation}
while for the Ellipse ($b<1$) it is:
\begin{equation}
\tau_{avg}^{Ellipse}=b~\tau_{min}+\frac{\pi}{4}~b~(\tau_{max}-\tau_{min}), 
\end{equation}
and for $b \gtrsim 1$: 
\begin{equation}
\tau_{avg}^{Ellipse}=\tau_{min}+\frac{\tau_{max}}{b}~\left(\frac{\sqrt{b^2-1}}{2}+\frac{b^2}{2}~sin^{-1}\frac{1}{b}\right), 
\end{equation}
and for the Gaussian distribution
\begin{equation}
\tau_{avg}^{Gauss}=1.251~\sigma~(\tau_{max}-\tau_{min})~{\rm erf}(0.707/\sigma)+\tau_{min}, 
\end{equation}
where ``erf'' is the Error Function. 
For the HPC, the spatially averaged optical depth 
is simply:
\begin{equation}
 \tau_{avg}^{HPC}=C_f~\tau^{HPC},
\end{equation} 
where $\tau^{HPC}$ is the same as $\tau_s$ in Eq. (6). 
We plot contour levels, assuming different IPC models, of 
$\tau_{avg}^{IPC}/\tau_{avg}^{HPC}$ 
in Figures 4, 5, and 6 for a set of $a$'s, $b$'s, and $\sigma$'s, with 
$\tau_{max}$ taking values between 0.5 and 15 and $\tau_{min} \le \tau_{max}$. 
We do not plot values for $\tau_{min} > 5$ because the IPC distributions 
would describe an opaque slab with complete coverage (or opaque 
partial coverage for the $b < 1$ Ellipse distribution). 
The result would be nearly saturated absorption troughs, 
and the method for deriving optical depths from the ratio of 
these troughs would fail.  
 
Notice that a small powerlaw index mimics complete coverage while a 
large one mimics partial coverage (Fig. 1). For the Ellipse and the 
Gaussian distributions the situation with $b$ and $\sigma$, respectively, 
is reversed. (Small values of $b$ and $\sigma$ simulate partial coverage, 
Figs. 2 and 3). A quick glance at Figures 4, 5, and 6 shows that the 
column densities implied by the HPC doublet analysis are comparable 
to the average column densities in the IPC distributions used to 
synthesize the ``observed'' lines. Notice that, in all cases, the HPC 
analysis returns precisely the optical depths values of the input IPC 
distribution  
when $\tau_{max}=\tau_{min}$. Therefore, the dashed diagonal line 
corresponds to the ratio being $\tau^{IPC}_{avg}/\tau^{HPC}_{avg}=1.0$. 
The HPC models also asymptotically approach the IPC case for small values 
of $a$ and large values of $b$ and $\sigma$, as is apparent from Figures. 
1, 2, and 3. In these cases, all of the IPC models actually resemble 
complete homogeneous coverage. An interesting situation occurs for the 
Ellipse distribution. When the semi-minor axis is less than 1, this HPC 
model becomes very similar to the IPC, especially for 
$\tau_{min} \approx \tau_{max}$ (Figs. 2 and 5). 

The contours in Figures 4, 5, and 6 run more or less parallel to the 
dashed diagonal line. Qualitatively, why do the contours behave the way they do? To answer this 
question, choose any IPC model and pick the plot corresponding to some specific $a$, $b$, or $\sigma$, 
as the case might be. At a particular $\tau_{max}$, as $\tau_{min}$ increases the distribution 
shape (Figs. $1-3$) begins more and more to resemble that of a homogeneously, completely covering 
cloud for the Gaussian and powerlaw distributions, or a homogeneous partial coverage for the Ellipse. 
In either case, in terms of average optical depths, the HPC output is nearly identical to the IPC input, 
and hence the ratio decreases from its maximum value when $\tau_{min}=0$  to unity as $\tau_{min}$ 
approachs $\tau_{max}$. This decrease is most rapid for peaked optical depth distributions, and hence the 
closely spaced contours. Sharply peaked distributions also lead to the largest  
differences between the ionic column densities in the IPC distribution and those derived under the HPC 
assumption. The decrease in ratio is less dramatic for optical depth distributions that more uniformly 
cover the source, and hence lead to contours that are close to unity, such as all the Ellipse clouds, 
the large $\sigma$ Gaussians, and the small 
$a$ powerlaws. As a matter of fact, the Ellipse clouds are always very similar to the HPC.  

The HPC results differ most dramatically from the IPC input when the IPC distribution 
is sharply peaked (large $a$ for the powerlaw or small $\sigma$ for the 
Gaussian) and there is a transition across this distribution from very
optically thick ($\tau_{max} >> 1$) to optically thin 
($\tau_{min} \approx 0$). For example, the parameters $a=10$, 
$\tau_{min}=0$, and $\tau_{max}=15$ in the powerlaw distribution (Eq. 11) 
lead to a spatially-averaged optical depth of 
$\tau_{avg}^{Powerlaw}=1.4$. Also $\sim$24\% of the source is covered by 
regions having $\tau>1$. An HPC analysis applied to the 
absorption doublets generated by this particular IPC distribution 
yeilds a $C_f=0.29$, which, as 
expected, is similar to the $\sim$24\% of the source covered by 
$\tau>1$ gas in the actual distribution. The optical depths
inferred from the HPC analysis are $\tau^{HPC}=2.9$ for lines of sight 
hitting the absorber, and zero elsewhere, implying a spatially-averaged 
optical depth of $\tau_{avg}^{HPC}=0.84$. We show in Figs. 7, 8, and 
9 the results of similar calculations for the powerlaw, Ellipse, 
and Gaussian distributions, respectively. The horizontal lines are 
the values of the corresponding $\tau_{avg}^{IPC}$ and $\tau_{avg}^{HPC}$.

It is not surprising that $\tau_{avg}^{HPC}$ in the 
previous paragraph is less than $\tau_{avg}^{Powerlaw}$ in the actual 
powerlaw distribution. The reason is that the absorption line 
strength is not sensitive to the exact value of $\tau$ in regions where 
$\tau$ is already $>>1$, e.g., on the right-hand side of Fig. 1a. For 
example, if we keep the same $\tau(x)$ distribution but truncate 
the peak at $\tau=10$ or even $\tau=5$, we would lower the value of 
$\tau_{avg}^{Powerlaw}$ without significantly affecting the 
absorption line strengths. Similarily, adding material (to increase 
the optical depth) in regions where $\tau$ is already $>>1$ could 
increase $\tau_{avg}^{Powerlaw}$ arbitrarily, but there would again 
be now change in the amount of absorption as registered by the 
absorption lines. 

The HPC analysis is remarkably accurate when applied to 
optical depth distributions that do not contain sharp spikes, i.e., 
having areas with $\tau>>1$, because the absorption lines themselves 
are not sensitive to optical depths in the $\tau>>1$ regions. Without 
specific knowledge of the true shape of the optical depth distribution, 
there is no prospect of using measured absorption lines to estimate 
the amount of material (column density) that might be present in 
very optically thick regions. However tighter constraints could 
be obtained by using, not just 
doublets with $2:1$ ratios, but multiplets and lines of more/less 
abundant ions that span a much wider range in $\tau$.  

\section{Applications}

The IPC models can also be applied to observed absorption line doublets 
(or multiplets) to derive key parameters of the optical depth 
distribution. To illustrate this, we use the HPC model, Eqs. (3) and (4) given 
$C_f$ and $\tau_s$, to simulate ``observed'' residual intensities, 
$I_s$ and $I_w$,  
in doublet lines whose optical depths are in a $2 : 1$ ratio. Basically, 
we follow the reverse approach of the previous section: we now assume 
that the doublet created with the HPC model is the ``observed'' and 
we solve for $\tau_{max}$ and $a$, $b$, or $\sigma$. (We fix 
$\tau_{min}=0$, with a little loss of generality, to reduce the 
number of unknowns in our equations). For every 
functional form of IPC, we get two equations with two unknowns 
(see Eqs. (14), (18), (19), and (23)), which we solve  
numerically using Newton's method (Press et al. 1992). The 
success of this approach hinges on providing a good initial guess. 
We find that trial and error guesses at $\tau_{max}$ and the shape 
parameter, guided by the input HPC values $C_f$ and $\tau^{HPC}$, 
lead more quickly to the best solution. The implementation 
of Newton's method also allowed us to confirm that the solutions arrived at 
are unique and not local minima which the numerical 
code converged to by mistake. 

The aim behind this exercise is to prove the feasibility of our 
IPC models in solving simultaneously, just as in the HPC case, 
for the optical depth and the ``shape'' parameter. We also 
note paranthetically that one can arrive back at the HPC input 
values if one uses the derived IPC parameters to simulate the residual 
intensities and then apply the HPC doublet analysis. 
This shows that the IPC doublet analysis we present in the 
this section is reliable. 
We test our method for a range of HPC parameters that would 
reveal the representative 
behavior of the solutions. Table 1 lists combinations of 
$C_f$ and $\tau^{HPC}$, used as input, and corresponding 
derived values of $\tau^{IPC}_{max}$, $a$, $b$, and $\sigma$. We chose 
three values of $\tau^{HPC}=0.5,1.0,$ and $5.0$, to span the range 
from optically thin to optically thick. 
The average optical depths are be calculated using Eqs. (25), 
(26), (27), (28), and (29). 
For $C_f$ we chose 0.1, 0.5, and 0.9 to investigate the effects 
of the degree of partial coverage on the resulting IPC solutions.

We compare the various quantities in Table 1. 
It is interesting to see how very low coverage leads to very large 
powerlaw indices and large derived values of $\tau_{max}$. 
For low optical depths with HPC the resulting maximum optical depth 
derived under the powerlaw IPC is about a factor of two larger. But the 
optically thick case ($\tau^{HPC}=5.0$) with low coverage ($C_f=0.1$) is 
much more extreme. We can understand this behavior because the 
powerlaw distribution describes complete coverage. It therefore 
strains to mimic a partial coverge situation with large $a$ values, 
resulting in a large spike in $\tau(x)$. This spike dominates 
$\tau_{avg}$, but it is purely an artifact of the functional form. 
It shows that the powerlaw distribution IPC is completely unreliable 
in this regime. One can always truncate the maximum optical depth value 
at some very high value without affecting strength of the resulting 
absorption lines. 

A quick glance at the Ellipse solutions, both in terms of 
$\tau_{max}$ and $\tau_{avg}$, shows 
that this IPC model is very similar to HPC (cf. previous section). 
On the other hand, the Gaussian distribution behaves like a combination 
of the the powerlaw, in terms of the peakedness, and the 
Ellipse. The Gaussian, like the powerlaw, is based on complete 
coverage. But the rapid decline in the Gaussian distribution 
from the core to the wings (Fig. 3) is better able to mimic 
partial coverage. Nonetheless, like the powerlaw case, optically 
thick clouds with small coverage fractions can lead to 
solutions with artificially large spikes in the Gaussian 
$\tau(x)$. We see that the derived average optical depths with the 
powerlaw, Ellipse, and Gaussian distributions are comparable 
to the average optical depth used to generate the HPC input, 
with the exception of the optically thick case $\tau^{HPC} =5.0$. 

The doublet analysis provides a shape parameter and $\tau_{max}$. 
But we are forced to assume a functional form that substantially 
affects the results (Table  1). Working with multiplets 
with ``n'' lines, or any combination of ``n'' lines whose 
$\tau$'s can be reliably tied together, would provide 
``n'' equations and ``n'' unknowns and therefore ``n'' 
constraints on the two-dimensional optical depth 
distribution. Such complicated procedure is, however, 
beyond the scope of this paper.  

\section{Two-Dimensional Optical Depth Distributions}
We describe in this section a procedure to re-parameterize an arbitrary 
two-dimenionsional optical depth distribution, $\tau(x,y)$, into a 
one-dimensional function $\tau(x)$. We show that the simple functional 
forms discussed above are relatively 
successful at representing more realistic and complicated situations.
Our starting point is to create an arbitrary 2D optical depth 
distribution. This 2D distribution, for the sake of definiteness 
but with some loss of generality, is the surface 
made from a collection of 36 overlapping Gaussian functions 
scattered over a plane (Fig. 10). The positions and heights 
of the Gaussian functions are chosen randomly, with numbers between 
0 and 1 for the abscissae/ordinates and between 0 and 3 for 
the heights (peak optical depths for the individual Gaussians). 
We pick the same 
Gaussian width, $\sigma =0.06$, for all the functions so as 
not to get very narrow or very wide spatial profiles. 
The outcome of this process 
is a 2D optical depth distribution across the surface of 
an emitter, which we assume to be emitting uniformly. 
The maximum optical depth in this distribution is 
$\tau_{max}=6.42$, the minimum optical depth is 
$\tau_{min}=0.0018$, and the average is $\tau_{avg}=1.17$. 

We then calculate the fraction of the area covered by 
optical depths between $\tau$ and $\tau+d\tau$. We 
divide the surface into a grid of 2D cells and 
count the number of cells with optical 
depths between $\tau$ and $\tau+d\tau$. This 
gives us $f^{2D}(\tau, \lambda)$ for our 2D distribution. 
It is now straightforward to integrate 
this function (recall $dx=(f(\tau,\lambda)d\tau)$) 
to get the equivalent 1D optical depth distribution
$\tau(x)$. We show the results in Figure. 11 (solid line). 

To calculate the residual doublet (with $2:1$ ratio) 
line intensities from such a distribution we numerically 
substitute $f^{2D}(\tau, \lambda)$ in Eq. (10) and 
evaluate the integral. This gives us two ``observed'' 
residual intensities for the doublet lines. We 
then apply our HPC and IPC analyses to this ``observed''
doublet to solve for the shape parameters and $\tau_{max}$, 
putting $\tau_{min}=0$ in the IPC case. We 
show the results in Table 2. The average optical depths 
calculated with all of the HPC and IPC models are 
comparable to the average optical depth in the 
original 2D distribution. This shows that the 
relatively simple functional forms we discussed in 
this paper provide practical representations to 
describe a complicated optical depth distribution.

We plot the corresponding 
HPC and IPC optical depth distributions in Figure 11 
for comparison with the equivalent 1D distribution. 
It is easy to see that the powerlaw and 
Gaussian functional forms are the closest to the  
re-parameterized 1D distribution, which was created 
from overlapping Gaussians. This re-parameterized 
1D function is more sharply peaked than the 1D IPC 
powerlaw that we derive from the doublet analysis. 
This shows that spikes in $\tau(x)$ can arise 
naturally from ``real'' 2D distributions, even if those 
distributions are based on a very simple, non-spikey 
function (a Gaussian). The ``spikeness'' in the 
1D re-parameterization (Fig. 11) is the result of having few 
peaks in the 2D distribution (Fig. 10).

\section{Conclusions}
Understanding the effects of the spatial distribution of ions 
in an absorbing region is essential to deciphering the 
physical and chemical states of the absorber. 
Our investigations of the effects of the spatial distribution of ions in an 
absorbing region on the derived column densities reveal a number of 
interesting 
properties. So long as an absorbing region is not optically thick 
we find that the average optical 
depth of the input distribution, i.e., the one used to generate 
the ``observed'' residual intensities in an absorption 
doublet, will lead to a comparable, to within a factor of less than 
1.5, average optical depth 
derived from a doublet analysis assuming a different optical 
depth distribution. Figures 4, 5, and 6 along with Tables 1 
and 2 atest to this. In Figures 4, 5, and 6 we start with 
an IPC distribution, $\tau_{max}$, $\tau_{min}$, and the appropriate 
shape parameter, and then derive $\tau^{HPC}$ and $C_f$ for the HPC case. 
We plot the ratio of the average optical depths, 
$\tau_{avg}^{IPC}/\tau_{avg}^{HPC}$, as contours, 
whose values are always less than $\sim 1.6$ in the part of parameter 
space we study. 

Table 1 shows the results of the reverse approach.  
We start by simulating an absorption doublet using an HPC distribution 
as input and use an IPC distribution to calculate the attributes 
of the absorbing region. It is easy to see that 
$\tau_{avg}^{HPC}\approx \tau_{avg}^{IPC}$, except  in the 
optically thick regime. We also should point out that we can 
get back to the initial HPC parameters from the IPC solutions by 
following the same approach employed of Figures 5, 6, and 7. 
This symmetry proves that our IPC doublet analysis method 
is reliable. We have extended the work of de Kool et al. (2002)
to show how, given a functional form in the IPC analysis, 
we can use observed lines (e.g., doublets) to infer key 
parameters about the inhomogeneous optical depth and 
column density distributions. 

We go a step further in \S 5 (Table 2). We start with an 
arbitrary 2D optical depth distribution and calculate the resulting doublet 
residual intensities, which we then analyze with the HPC and IPC 
models. The last column in Table 2 shows that the resulting average 
optical depths are almost identical to the average optical 
depth of the initial 2D distribution. All this discussion 
leads us to an important conclusion: Our investigation shows 
that the spatial distribution of ions has a minimal effect on the 
computed average column density. Given the fact that spikes could arise 
from not so exotic 2D distributions (Figs. 10 and 11) and 
that the re-parameterized 1D function is closest to a 
powerlaw (Fig. 11), we conclude that the IPC powerlaw doublet 
analysis should yield more representative parameters 
of the absorbing line region. Solving Eq. (14) for a doublet that 
provides two equations 
in two unknowns ($\tau_{max}$, $a$, with $\tau_{min}=0$), 
is relatively easy. In a future paper we will apply our 
methods to observed quasar absorption lines and present the numerical 
codes that implement our IPC doublet analysis.  

\noindent {\it Acknowledgements:} We wish to acknowledge support 
through NSF grant AST99-84040.

\begin{figure}
\centerline{
\psfig{figure=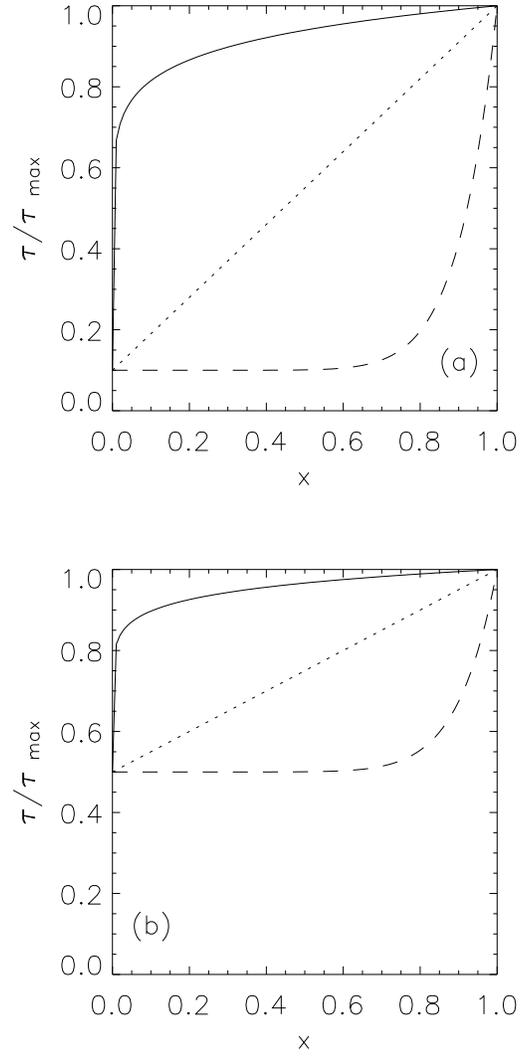}
}
\caption{Powerlaw distribution with $\tau_{min}=0.1~\tau_{max}$ $(a)$ 
and $\tau_{min}=0.5~\tau_{max}$ $(b)$. Solid, dotted, and 
dashed lines for $a=0.1, 1.0,$ and 10, respectively.} 
\end{figure}

\begin{figure}
\centerline{
\psfig{figure=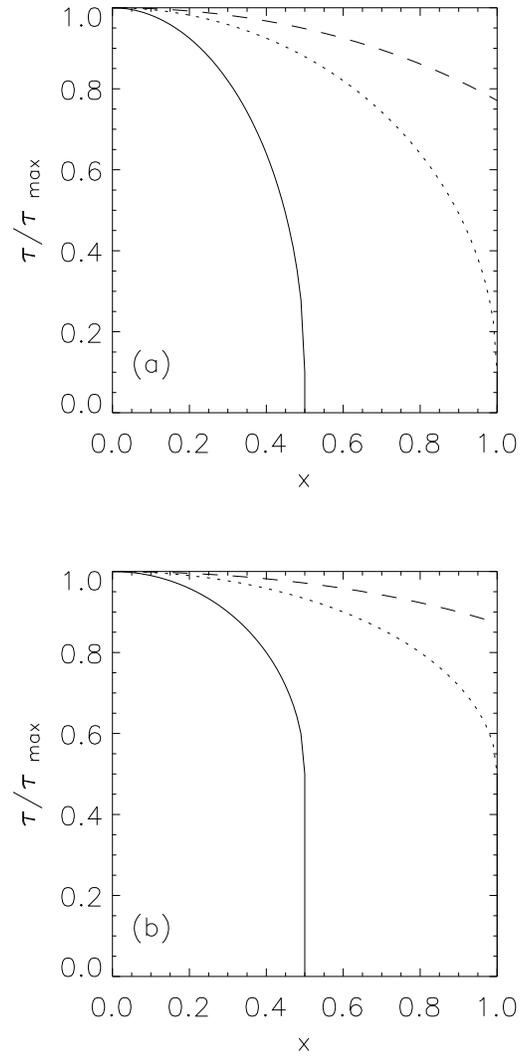}
}
\caption{Ellipse distribution with $\tau_{min}=0.1~\tau_{max}$ $(a)$ 
and $\tau_{min}=0.5~\tau_{max}$ $(b)$. Solid, dotted, and 
dashed lines for $b=0.5, 1.0,$ and 1.5, respectively.} 
\end{figure}

\begin{figure}
\centerline{
\psfig{figure=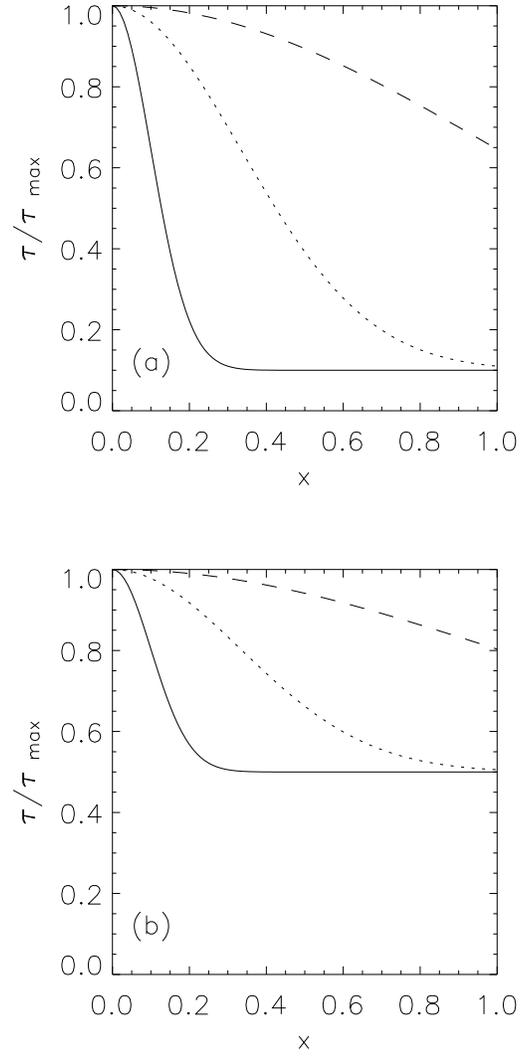}
}
\caption{Gaussian distribution with $\tau_{min}=0.1~\tau_{max}$ $(a)$ 
and $\tau_{min}=0.5~\tau_{max}$ $(b)$. Solid, dotted, and 
dashed lines for $\sigma=0.1, 0.33,$ and 1.0, respectively.} 
\end{figure}

\begin{figure}
\centerline{
\psfig{figure=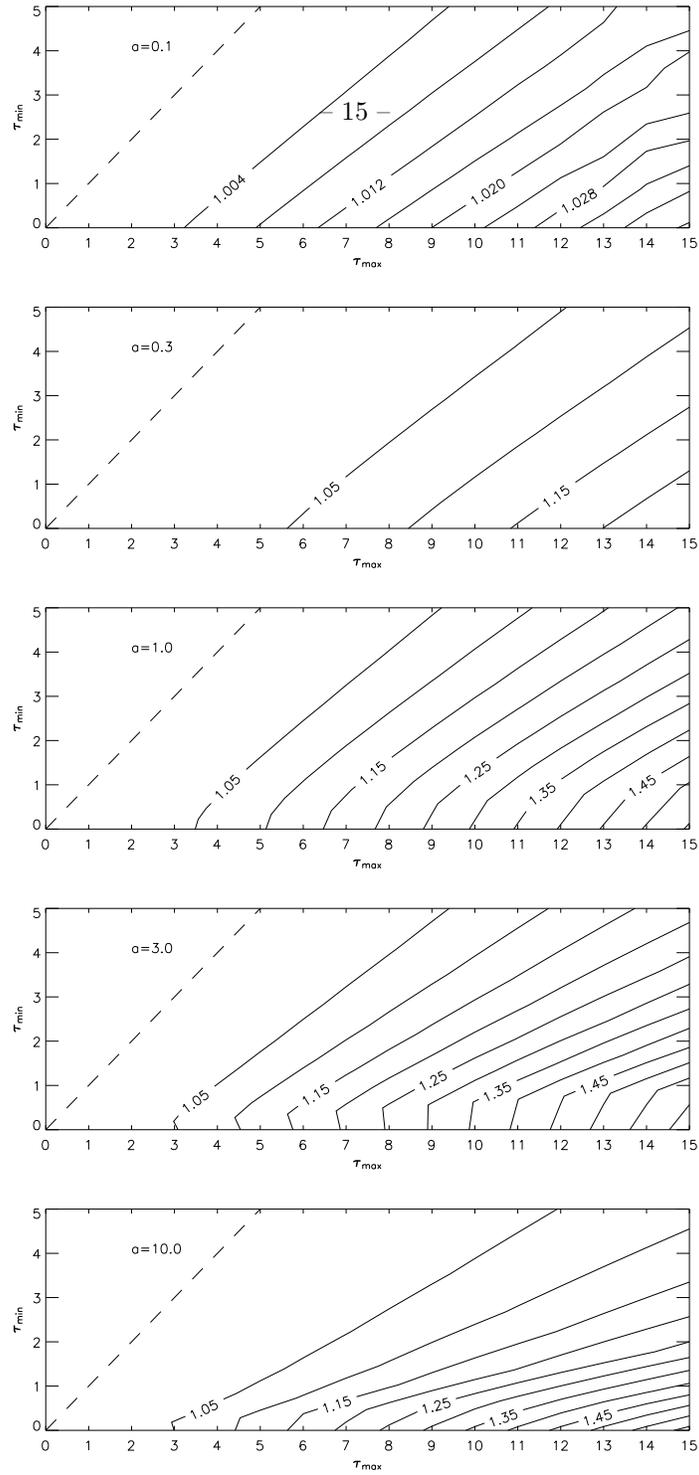}
}
\caption{Powerlaw Distribution vs. HPC. Contours show the ratio of 
spatially-averaged optical depths calculated in the IPC model to that 
computed under HPC assumptions, i.e., $\tau_{avg}^{IPC}/\tau_{avg}^{HPC}$. 
The dashed curve indicates $\tau_{min}=\tau_{max}$, which also 
corresponds to $\tau_{avg}^{IPC}/\tau_{avg}^{HPC}=1$, because in this 
limit, the $\tau(x)$ distribution (Eq. 11) is homogeneous.}
\end{figure}

\begin{figure}
\centerline{
\psfig{figure=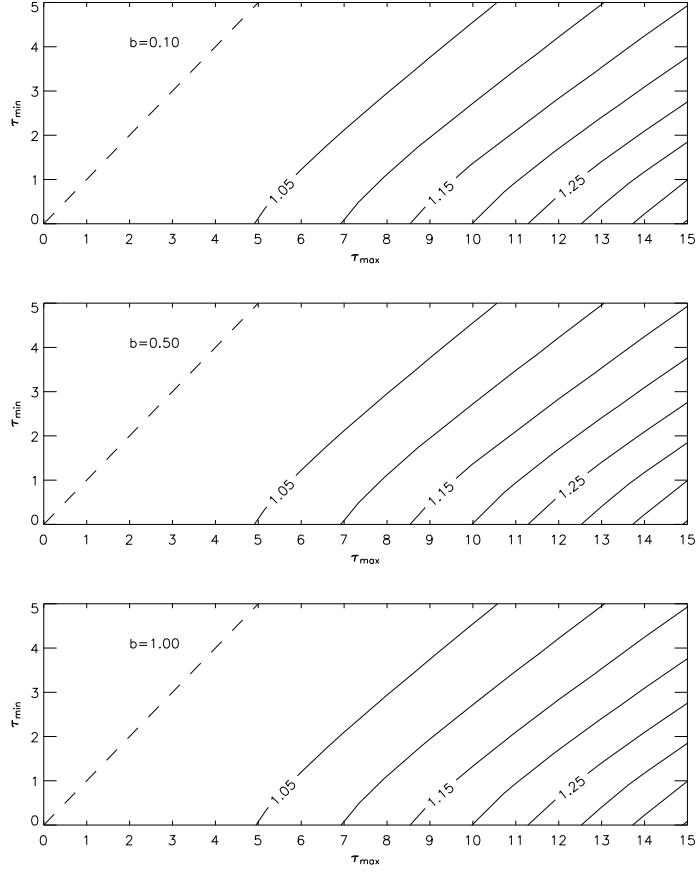}
}
\caption{Same as Fig. 4 but for Ellipse Distribution vs. HPC. For 
$b > 1.0$, IPC model is the same as homogeneous, complete coverage, and 
hence the ratio of optical depths is always 1.}
\end{figure}

\begin{figure}
\centerline{
\psfig{figure=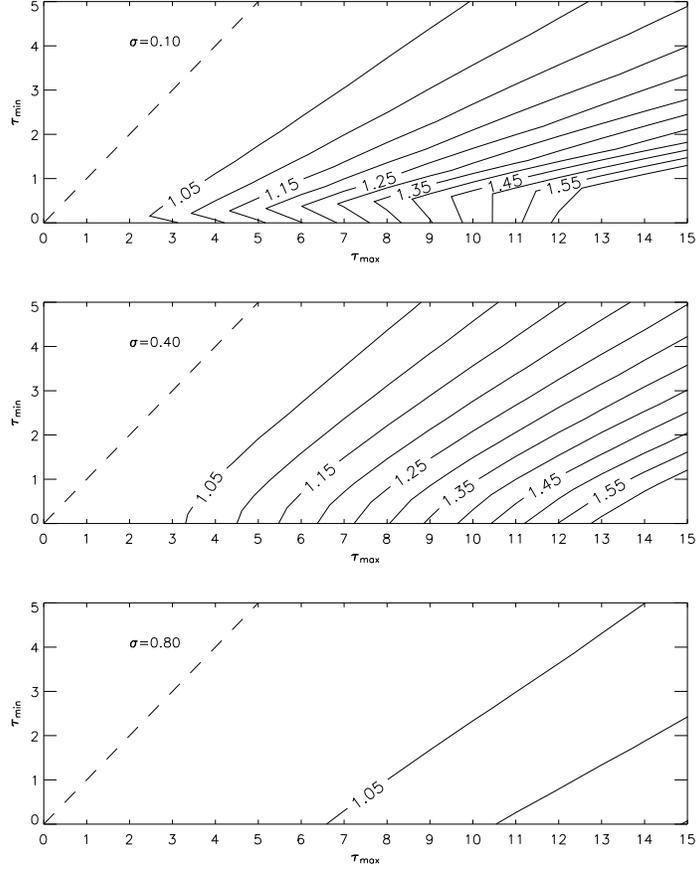}
}
\caption{Same as Fig. 4 but for Gaussian Distribution vs. HPC.
For $\sigma > 0.8$, IPC model is the same as homogeneous, complete 
coverage, and hence 
the ratio of optical depths is always 1.}
\end{figure}

\begin{figure}
\centerline{
\psfig{figure=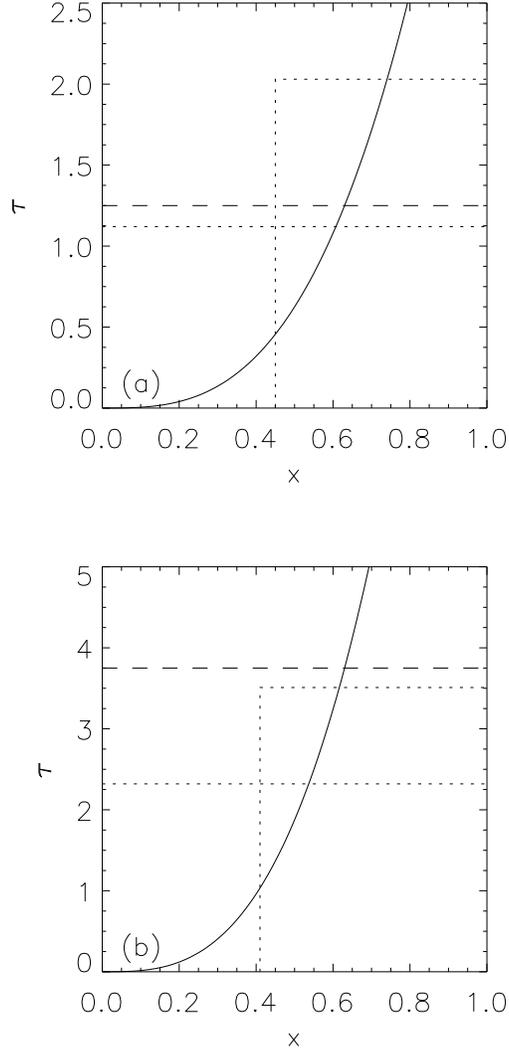}
}
\caption{Powerlaw distribution for $a=3$ (solid curves) 
with $\tau_{max}=5$ (top panel) and $\tau_{max}=15$ (bottom). The 
dashed horizontal lines show the average optical depth, 
$\tau_{avg}^{IPC}$,  of these distributions. The dotted rectangles show 
the HPC distributions of the optical depths, as derived from the 
doublet analysis using the IPC distributions as input (see \S 3). 
The horizontal dash-dotted lines mark the values of $\tau_{avg}^{HPC}$, 
the average optical depths of the HPC distributions. 
The plots were truncated at $\tau < \tau_{max}$ 
to aid in noticing the differences between the average optical depths.}
\end{figure}

\begin{figure}
\centerline{
\psfig{figure=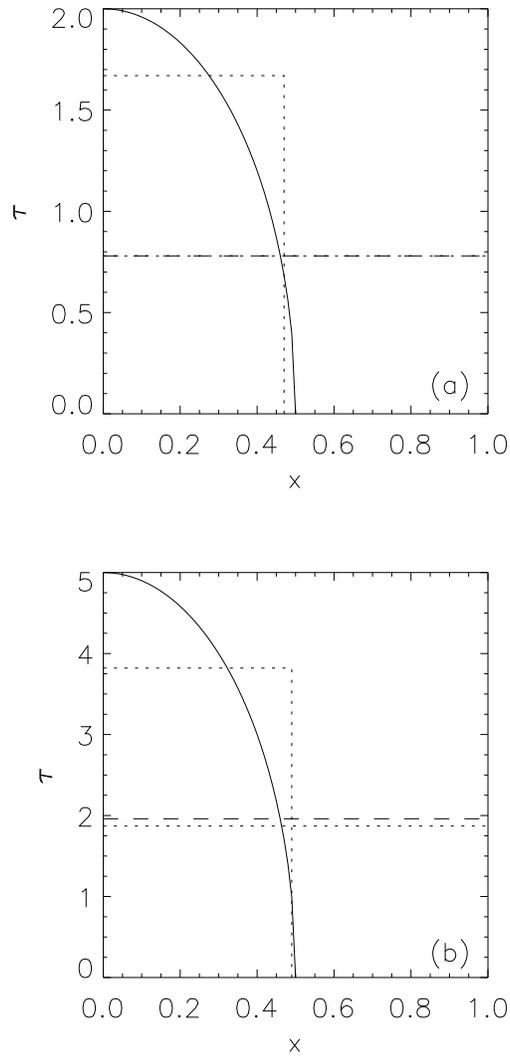}
}
\caption{Same as for Fig. 7 but for an Ellipse distribution ($b=0.5$) 
with $\tau_{max}=2$ (top, a), and $\tau_{max}=5$ (bottom, b). We 
chose lower optical depth values to avoid black troughs.}
\end{figure}

\begin{figure}
\centerline{
\psfig{figure=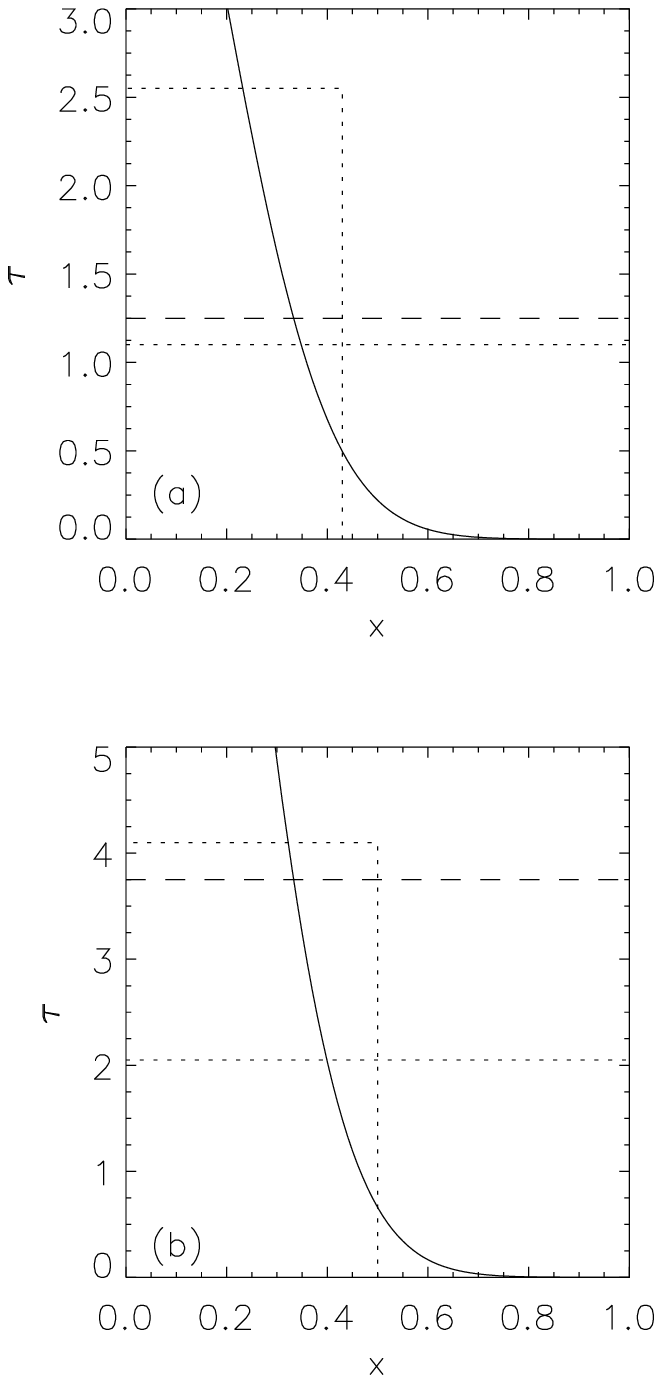}
}
\caption{Same as for Fig. 7 but for an Gaussian distriubution 
with $\sigma=0.2$ with $\tau_{max}=5$ (top, a), and 
$\tau_{max}=15$ (bottom, b).}
\end{figure}

\begin{figure}
\centerline{
\psfig{figure=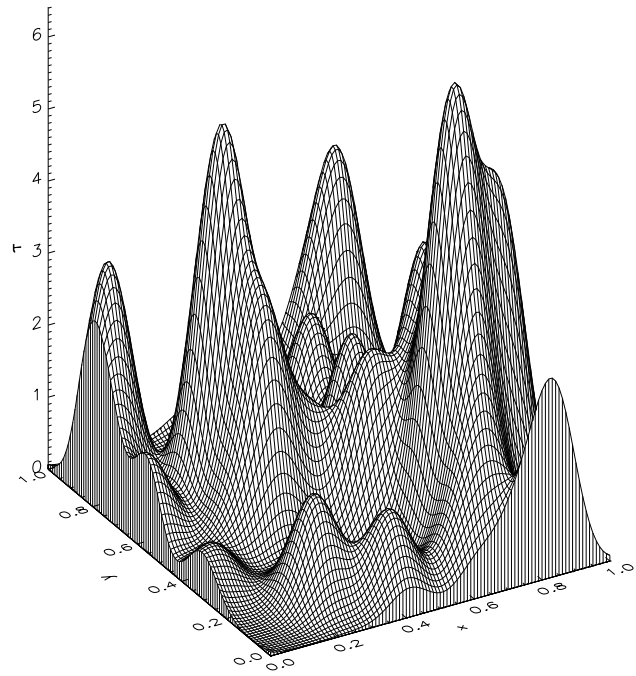}
}
\caption{Two dimensional optical depth distribution synthesized from 36 overlapping 
Gaussians.}
\end{figure}

\begin{figure}
\centerline{
\psfig{figure=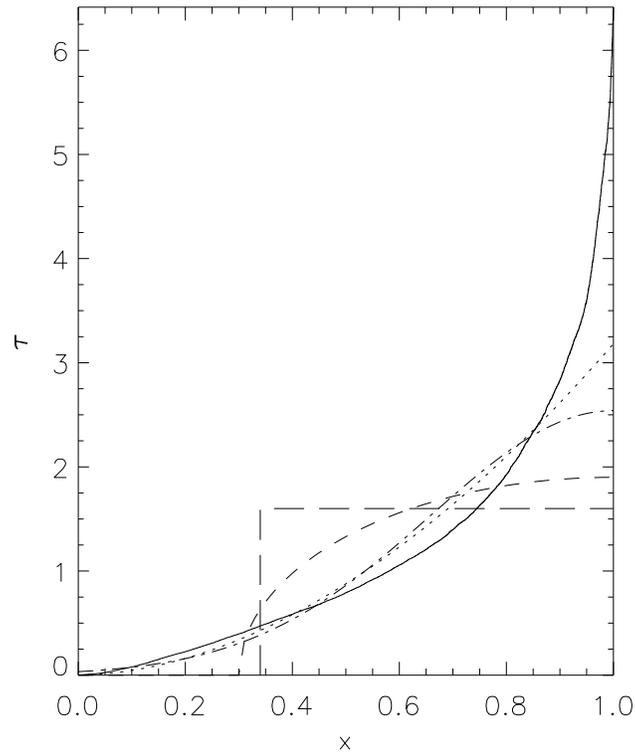}
}
\caption{The 1D re-parameterization (solid curve) of the 2D optical 
depth distribution shown in Figure 10 is shown with the powerlaw 
(dotted), Ellipse (dashed), Gaussian (dash-dotted), and HPC 
(long dashed) distributions derived for the doublet analysis 
in (\S 5).}
\end{figure}

\begin{table}
\begin{center}
\title{\small \rm Table 1: IPC Solutions}
\vspace*{0.1in}
\begin{tabular}{cccc}
\tableline
\tableline
$C_f, \tau^{HPC}, \tau_{avg}^{HPC}$ & $a, \tau^{Powerlaw}_{max}, \tau_{avg}^{Powerlaw}$ & $b, \tau^{Ellipse}_{max}, \tau_{avg}^{Ellipse}$ & $\sigma, \tau^{Gaussian}_{max}, \tau_{avg}^{Gaussian}$ \\
\tableline
0.1, 0.5, 0.05	&20.1, 1.1, 0.05&0.14, 0.4, 0.04&0.06, 0.67, 0.05	\\
0.1, 1.0, 0.10	&22.1, 2.4, 0.10&0.10, 1.1, 0.09&0.05, 1.43, 0.09	\\
0.1, 5.0, 0.50	&832, 3415, 4.10&0.10, 6.9, 0.54&0.04, 36.4, 1.82	\\
0.5, 0.5, 0.25	&2.6, 0.9, 0.25	&0.70, 0.4, 0.22&0.29, 0.65, 0.24	\\
0.5, 1.0, 0.50	&2.9, 2.0, 0.51	&0.55, 1.2, 0.52&0.27, 1.43, 0.48	\\
0.5, 5.0, 2.50	&9.6, 444, 41.8	&0.51, 6.9, 2.76&0.18, 36.4, 8.19	\\
0.9, 0.5, 0.45	&0.5, 0.7, 0.47	&1.18, 0.5, 0.46&0.54, 0.69, 0.44	\\
0.9, 1.0, 0.90	&0.6, 1.4, 0.87	&1.00, 1.1, 0.86&0.66, 1.19, 0.85 	\\
0.9, 5.0, 4.50	&1.4, 20.4, 8.5	&0.91, 6.9, 4.90&0.33, 36.4, 15.0	\\
\tableline
\end{tabular}
\end{center}
\end{table}

\begin{table}
\begin{center}
\title{\small \rm Table 2: 2D Distribution \& HPC/IPC Models}
\vspace*{0.1in}
\begin{tabular}{cccc}
\tableline
\tableline
Model   & Geometry& $\tau_{max}$ & $\tau_{avg}$ \\
\tableline
2D      &  ---   &     6.42     &     1.17     \\
HPC	& 0.66    &     1.60	 &     1.06	\\
Powerlaw& 1.86    &     3.18	 &     1.11	\\
Ellipse & 0.70    &     1.90	 &     1.05	\\
Gaussian& 0.34    &     2.54	 &     1.08	\\

\tableline
\end{tabular}
\end{center}
\end{table}

\end{document}